\documentclass[twocolumn, letterpaper, pra, showpacs, aps]{revtex4-1}
\usepackage[english]{babel}
\usepackage[applemac]{inputenc}
\usepackage{pgf}
\usepackage{amsmath}
\usepackage{amsfonts}
\usepackage{mathtools}
\usepackage{upgreek}
\newcommand{\ie}{i.~e.,~} 
\newcommand{\eg}{e.~g.,}

\usepackage{hyperref}

\hypersetup{colorlinks, citecolor=blue, linkcolor=black, urlcolor=black}

\begin{document}
\title{A characteristic function approach to the Jaynes-Cummings model
  revivals}

\date{\today}

\author{Hudson Pimenta}
\author{Daniel F. V. James}
\affiliation{Department of Physics, University of Toronto, 60
  St. George Street, Toronto, Ontario M5S 1A7, Canada}

\begin{abstract}
A two-level atom interacting with an electromagnetic mode in a cavity
experiences atomic inversion collapses and revivals. They are an indirect signature of
the field quantization, and also hold information about the
mode. Thus, they may be harnessed for quantum-state reconstruction. In this work, we study the revival
structures with the characteristic function approach. The
characteristic function is essentially a spectral decomposition of the
photon-number probability distribution. Exploiting the
characteristic function periodicity, we find that the inversion can be understood as the result of interference
between a set of structures akin to a free quantum-mechanical
wave packet, each structure corresponding to a snapshot of this packet
for different degrees of dispersion. The packet Fourier representation determines the photon-number distribution of the electromagnetic
mode. We then derive an integral equation whose solution yields the underlying packets. This approach allows the
retrieval of the field photon-number distribution directly from the
inversion under fairly general conditions, and paves the way for
a new partial tomography technique.
\end{abstract}

\maketitle

\section{Introduction}
\label{sec:introduction}
Experiments in quantum optics have by now unequivocally established the
granular and quantum nature of the electromagnetic field
\cite{Clauser:1974hc,Grangier:1986ge,Hong:1987gm,Thorn:2004js}.
Technology has since improved to the point that many sorts of quantum
states of the field can be
synthesized \cite{WU:1986jo,Slusher:1985iw,Ourjoumtsev:2007wi,Hofheinz:2008dq,Hofheinz:2009ba}. They
can also be studied in much more controllable environments, such as
cavities, giving rise to a field called cavity electrodynamics (cavity
QED)
\cite{Meschede:1985ep,Walther:1988bj,Horoche:1989up,Walther:2006da}. In
a typical cavity experiment, an atom traverses a cavity and interacts with
an electromagnetic field mode. Understanding the dynamics of this
interaction enables further probing of the predictions of quantum
mechanics and its exploitation for various applications
\cite{Brune:1996bv,Hagley:1997fi,Brattke:2001ee}.

In the context of cavity QED, an important model is the Rabi model
\cite{Rabi:1936kp,Rabi:1937ck}, which describes a single
electromagnetic mode coupled to a two-level atom. Though it can be solved exactly \cite{Braak:2011hc}, this model affords a much
simpler picture when the field-atom coupling is weak. Then the model may be
approximated by the paradigmatic Jaynes-Cummings model (JCM) \cite{Jaynes:fa,Shore:1993gc}. Despite its
simplicity, the JCM has remarkable features associated with the
granular nature of photons. Some of the most striking of these features are the atomic inversion collapses and revivals \cite{Eberly:1980bw,Shore:1993gc}.

The collapses and revivals are a result of interference between inversion oscillations with different
frequencies, each corresponding to a possible number of photons $n$
inside the cavity. The
revivals, in particular, are only possible because the number of
photons is discrete. Therefore, they are an indirect signature of
the EM field quantization. Moreover,
the revivals quasi-periodicity provides evidence for the JCM anharmonic
energy ladder (which has been
observed through population measurements \cite{Brune:1996kh}
and, more directly, through spectroscopy
\cite{Fink:2008bl}). Inversion revivals have been extensively
investigated in the literature \cite{Eberly:1980bw, Narozhny:1981dx,
  Knight:1982kz, Puri:1986kp, Phoenix:1988dc, Gora:1994dp,Jonathan:1999uf}.

Besides their historical relevance, the revivals also hold
potential for characterizing the field state, since the
inversion profile is directly dictated by the field photon-number
amplitudes. Determining the field quantum state from experimental data is the aim of quantum-state
reconstruction \cite{LEONHARDT:1995wy, Weigert:2009jg}. In the more traditional
approach, we represent a quantum state by a density matrix, the
elements of which are to be determined by repeated
measurements of a set of observables \cite{Buzek:1996ke}. For example,
in the context of a two-level system, these observables may be the
ones associated to the Pauli matrices \cite{James:2001bb}. 

A mode of the electromagnetic field, however, is more complicated. Firstly, since it is a
quantum harmonic oscillator, there are infinitely many matrix elements to be
determined. Moreover, when its density operator is
represented in the basis of Fock states, population measurements
require photon-number resolving techniques, which have become
available only much more recently \cite{Schuster:2007ki, Guerlin:2007ca}.

For this reason, historically, tomography for the field took a
different path, by exploring representations of the state in
terms of phase space quasi-probability distributions \cite{Moyal:436529,Cahill:1969vq}. One of the most
popular distributions is the Wigner function
\cite{Wigner:1932cz}. It contains information about the generalized field
quadrature probability distributions
\cite{LEONHARDT:1995wy}. By measuring these distributions through balanced homodyne detection, it is possible to reconstruct
the Wigner function
\cite{VOGEL:1989tj,Smithey:1993er,Lvovsky:2009fr}. Later proposals use the same data to obtain the density matrix directly
\cite{dAriano:1994iy,dAriano:1995ia}. Unbalanced homodyning is also a
possibility \cite{Wallentowitz:1996fg,Opatrny:1997cs}.

Yet another tomographic approach consists of coupling the field to an auxiliary
simpler system from which information about the field is
retrieved indirectly. It is in this context that cavity QED and the
JCM revivals insert. It has been shown that, when an inversion revival
may be isolated, it may be used to retrieve photon-number distributions
\cite{Fleischhauer:1993dw}. A phase-sensitive scheme that uses population curves for different atomic
coherent superpositions for reconstruction has also been proposed
\cite{Bardroff:1995gd}. An alternative that avoids the coherent
superposition technicalities consists of displacing the field state instead
\cite{Bodendorf:1998fb}. Moreover, atomic population measurements may
also be used to probe the Wigner function \cite{Lutterbach:1997cn}.

With the goal of investigating alternative approaches for quantum harmonic
oscillators tomography, we investigate the Jaynes-Cummings inversion via the
approach of the field mode distribution characteristic function
\cite{Lukacs:108658}. Characteristic functions are the
Fourier representation of a distribution. In this paper, we use their properties to
decompose the inversion into into much simpler, localized in time,
packets, with shape dictated by the photon-number characteristic
function. We show that each packet is akin to a snapshot of a free
quantum-mechanical wave-packet at a different effective time. Hence, just as in quantum
mechanics, knowledge of a single packet is enough to
generate every other packet and, therefore, the whole inversion.

The advantages of this approach are as follows. Firstly, it shows
that the inversion contains highly redundant information (besides being an awkward function for analytical
and numerical manipulations due to its slowly-decreasing
behaviour). By contrast, a single one
of the packets we introduce in this work contains
the complete physics of the inversion. We also argue that, when they do not
overlap, each may be identified with a single revival, in which case
the photon-number distribution may be retrieved immediately \cite{Fleischhauer:1993dw}.

Secondly, even when revivals cannot be resolved, the underlying
picture of a superposition of packet persists. It is then that the
snapshot decomposition is the most advantageous: we use it in this work cast aside the
limitation of non-overlapping revivals, and retrieve the photon-number
distribution under much more general conditions. The key idea is that
the snapshot to be retrieved is usually concealed behind an overly
complex inversion profile. However, within a properly chosen time window, this
inversion is simply this snapshot, albeit contaminated by tails of adjacent
snapshots. 

If different snapshots were all unrelated, this would spell doom for any attempts at
its retrieval.
However, due to the
quantum-mechanical analogy, a sum of different snapshots may be
ultimately reduced to an integral equation involving just a single one
of them. Solving
this equation yields the packet with full information about the
inversion and, therefore, the photon-number distribution. In this
work, we illustrate our
approach mostly through coherent states due to their
simplicity. However, we emphasize the generality of this method, which
will be explored more meticulously, and for a larger variety of states, in a future
work. The present work lays the groundwork for this new form of partial tomography.

This paper is organized as follows.
Sec.~\ref{sec:jcmmodel} reviews the JCM and its population
quasi-periodic revivals. We also introduce periodic revivals, due to their simplicity and also because many of their
qualitative features persist in the quasi-periodic
case. Sec.~\ref{revivalcharacteristic} considers the periodic
and quasi-periodic revivals in
terms of the characteristic function. We use its properties to split
the revivals and reveal the packets underlying the
inversion for a field in a general state. We consider a coherent
state as an example, but reiterate that generality of this decomposition. Sec.~\ref{understandingcats} exemplifies the formalism
previously developed for a cat state. Finally, Sec.~\ref{overlapping} highlights
the novelty of this approach by discussing the extraction the
packets and the photon-distribution for a very general inversion profile. Finally,
Sec.~\ref{sec:summary} summarizes our results. 

\section{Collapses and revivals}
\label{sec:jcmmodel}
In this section, we review the collapses
and revivals in the JCM. The JCM descends from the Rabi model, which describes a single
electromagnetic mode coupled to a single two-level atom
\cite{Rabi:1936kp,Rabi:1937ck}. The Rabi Hamiltonian reads ($\hbar = 1$)
\begin{equation}
  \label{JCMHamiltonianRabi}
  \hat{H}_{R} = \omega \hat{a}^\dagger \hat{a} + \frac{\omega_0}{2} \hat{\sigma}_z + g\hat{\sigma}_x
  (\hat{a}^\dagger + \hat{a}),
\end{equation}
where $\hat{a}$ is the annihilation operator for a photon in an
electromagnetic mode of frequency $\omega$, $\omega_0$ is the
splitting between the two atomic levels and $g$ is field-atom coupling
constant. The two-level atom has been
mapped into a pseudospin $\hat{\mathbf{\sigma}}$, with ground state
$|g \rangle$ and excited state $|e \rangle$ corresponding
to spin down and up in the $z$-direction, respectively. In this language, $\hat{\sigma}_x =
\hat{\sigma}_+ + \hat{\sigma}_-$
represents the atomic dipole moment. Assuming weak field-atom coupling, \ie $g \ll \omega,~
\omega_0$, we neglect terms in the Hamiltonian
proportional to $\hat{a} \hat{\sigma}_-$ and $\hat{a}^\dagger
\hat{\sigma}_+$. They lead to
costly energy transitions ($\sim \omega + \omega_0$) when compared to those generated by
$a \sigma_+$ and $a^\dagger \sigma_-$ ($\sim \omega - \omega_0$). This
leads to
\begin{equation}
  \label{jcmrwa}
    \hat{H}_{JC} = \omega \hat{a}^\dagger \hat{a}  + \frac{\omega_0}{2} \hat{\sigma}_z + g( \hat{a}^\dagger \hat{\sigma}_- + \hat{a} \hat{\sigma}_+),
\end{equation}
which is known as the JCM Hamiltonian.

In this small $g$ limit, an atom transitioning away from the excited
state (ground state) is always followed by a photon emission
(absorption). Hence, the state of the system initially given by $ | e, n
\rangle$, where $| n \rangle$ is the Fock state with $n$ photons, will
oscillate between $| e, n \rangle$ and $|g, n+1 \rangle$ as
\begin{align}
  \label{eq:70}
  \hat{U}(t) | e, n \rangle = &\cos \left( \frac{\Omega_n }{2} t \right) |e, n \rangle \\
& + i \sin \left(\frac{\Omega_n}{2} t \right) |g, n+1 \rangle, \nonumber
\end{align}
where $\Omega_n = 2g \sqrt{n+1}$.

The atomic populations in the ground and excited states are then $\displaystyle P_g (t) = \sin^2
\left(\frac{\Omega_n}{2} t \right)$ and $\displaystyle P_e (t) = \cos^2 \left(
  \frac{\Omega_n }{2} t \right)$, respectively. It is customary to define the
population inversion as the difference between these populations: $W(t) \equiv P_e (t) - P_g (t)$. For $|e, n \rangle$ as initial state,
inversion is simply $\cos (2g \sqrt{n+1} t)$. However, more generally,
the field state
is a superposition of Fock states $|n \rangle$ with different
photon-number amplitudes $c_n$. Assuming that atom is still initially
excited, the inversion takes the more general form 
\begin{equation}
\label{atomicpopulation}
  W(t) = \sum_{n =0}^{+\infty} |c_n|^2 \cos \left( 2g \sqrt{n+1} t \right),
\end{equation}
where $|c_n |^2 \equiv P_n$ is the field photon-number distribution.

\begin{figure}[t]
\includegraphics[width=1.\columnwidth]{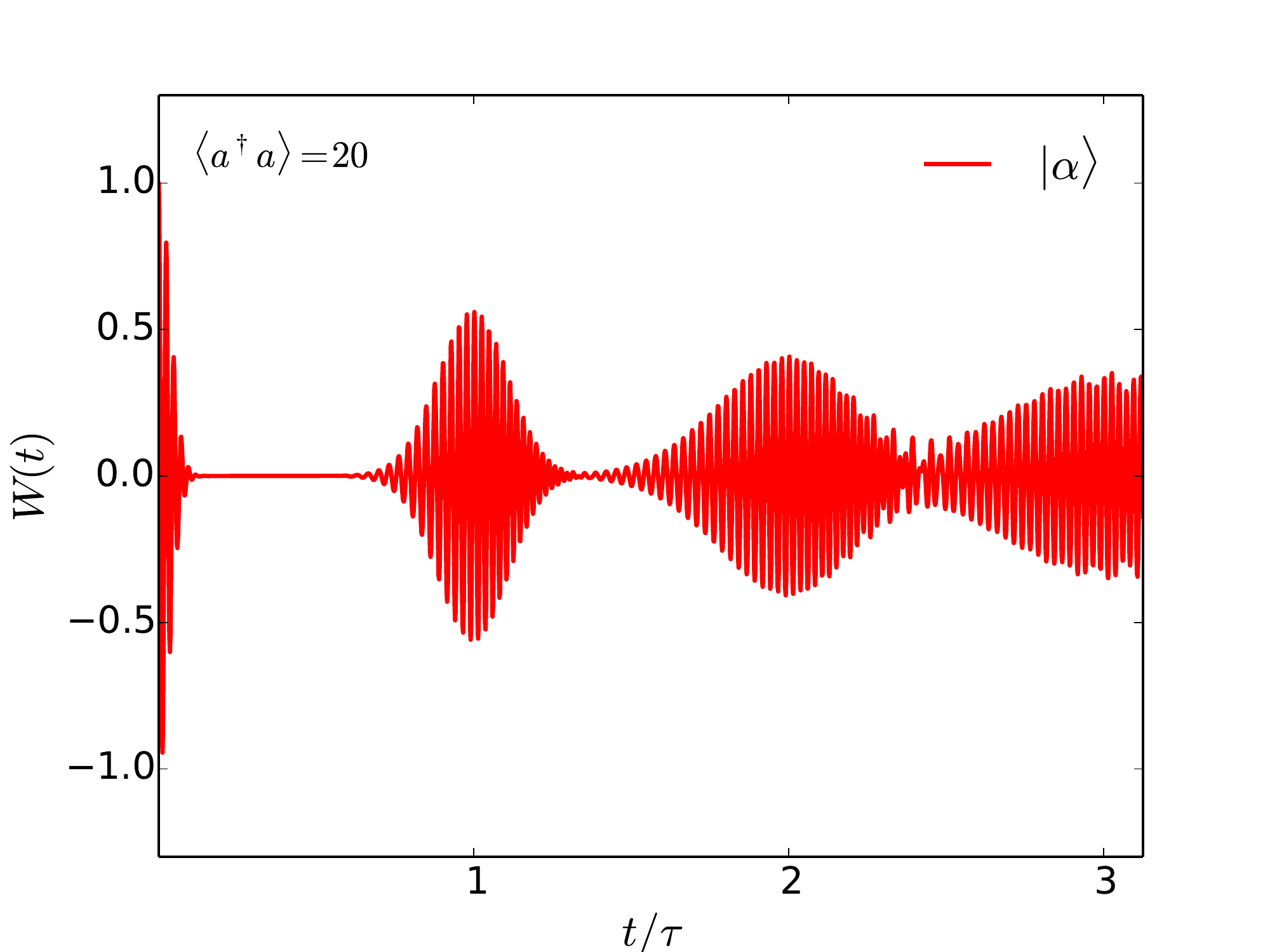}
\caption[revivalcoherent]{\label{revivalcoherent} Atomic inversion of Eq.~\eqref{atomicpopulation} as
a function of time. The atom is
initially excited and the field is in a coherent state with average
photon-number of $20$. Time is measured in
units of $\tau$, where $\tau$ is the time for which the first revival
amplitude is maximum. The dephasing of the oscillators in Eq.~\eqref{atomicpopulation} lead to inversion collapses. With discrete yet
incommensurable frequencies, the inversion has revivals, but they are
only quasi-periodic. The revival peaks are approximately equally
spaced by $\tau$.}
\end{figure}

A great deal of attention has been given to the inversion because it
provides evidence of the electromagnetic field quantization through
its collapses and revivals \cite{Shore:1993gc}. We illustrate the inversion for a coherent state in Fig.~\ref{revivalcoherent}. For short
times, $W(t)$ is dominated by Rabi-like oscillations. As
the oscillators of Eq.~\eqref{atomicpopulation}
dephase, they interfere destructively, causing the collapse. Still, the discreteness of the frequencies, a direct consequence of the field quantization, allows for
population revivals at later times. The revivals are
not, however, periodic, since some frequencies are incommensurable,
\ie their ratios are irrational numbers.  

The inversion $W(t)$ is interesting also because it holds information
about the field photon-number distribution $P_n$. However, the
incommensurable frequencies hampers the distribution retrieval: Eq.~\eqref{atomicpopulation} looks like a Fourier
series, but it is not, due to the frequencies incommensurability. An
inversion formula is known when a single
revival of $W(t)$ can be isolated
\cite{Fleischhauer:1993dw}. 

\begin{figure}[t]
\includegraphics[width=1.\columnwidth]{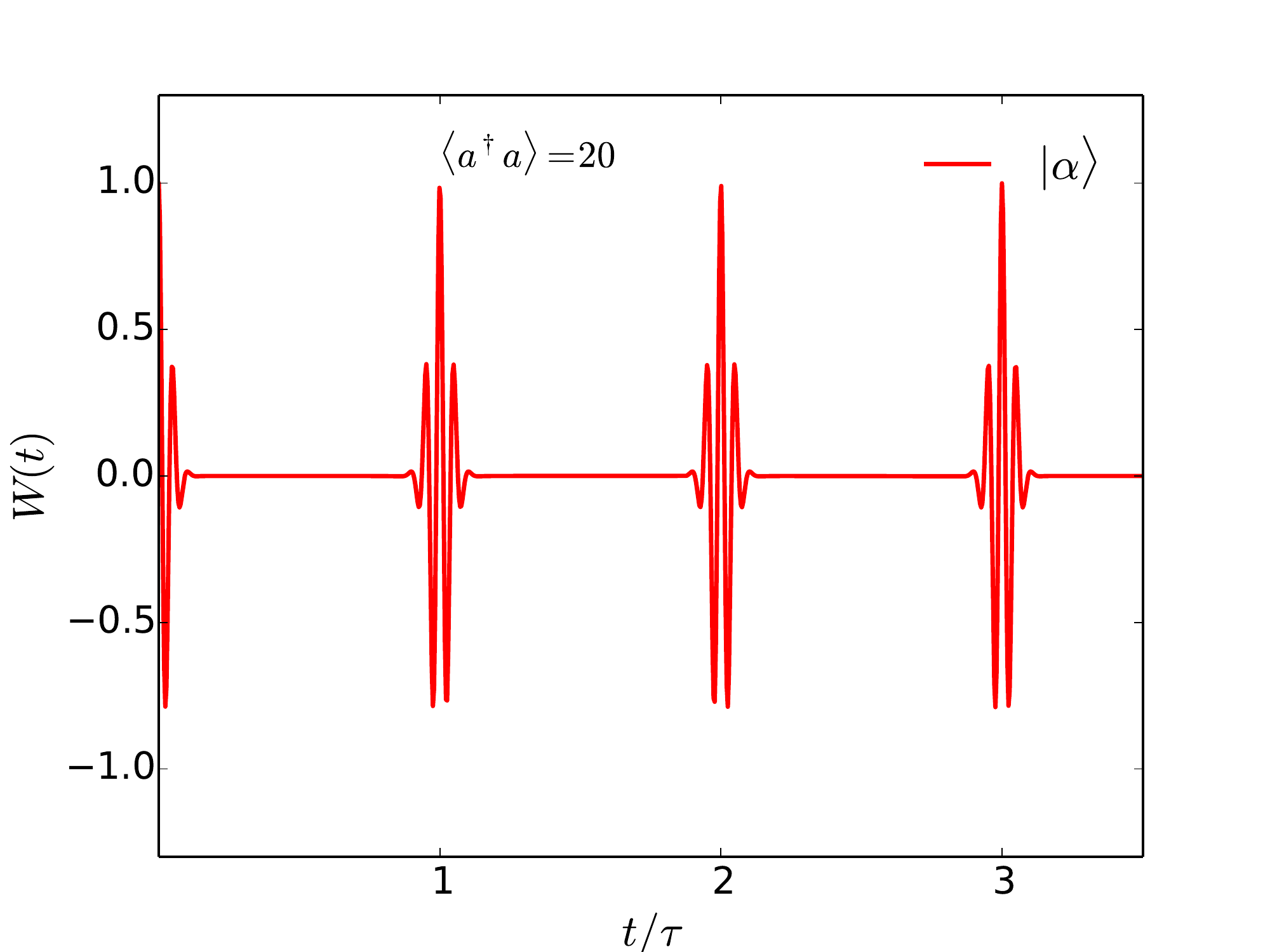}
\caption[revivalcoherent]{\label{revivalperiodiccoherent} Inversion as
a function of time when the frequencies are commensurable, as in
Eq.~\eqref{atomicpopulationnewmodel}. The atom is
initially excited and the field is in a coherent state with average
photon-number of $20$. Time is measured in units of
$\tau$, where $\tau = \pi$ is the inversion period. With only commensurable frequencies, the inversion
is exactly periodic.}
\end{figure}

To support the next section discussion, we also present (exactly)
periodic revivals. They may be seen as mathematical
constructs defined by the replacement $\sqrt{n+1} \to n$ in Eq.~\eqref{atomicpopulation}:
\begin{equation}
  \label{atomicpopulationnewmodel}
  W_p(t) = \sum_{n=0}^{+\infty} |c_n|^2 \cos \left( 2 g n t\right).
\end{equation}
The lower index $p$ is a reminder that this inversion is not the same
as the JCM inversion. The frequencies in Eq.~\eqref{atomicpopulationnewmodel} are all
commensurable. Hence $W_p (t)$ is exactly periodic, the period given by
$T = \pi / g$. Unlike Eq.~\eqref{atomicpopulation},
Eq.~\eqref{atomicpopulationnewmodel} is an actual Fourier series, and
may be inverted immediately. We illustrate the periodic revivals in
Fig.~\ref{revivalperiodiccoherent} under the same conditions of
Fig.~\ref{revivalcoherent}. The periodic revivals are very
useful for interpreting the JCM revivals, since both share many
qualitative features. For instance, in both
Figs. \ref{revivalcoherent} and \ref{revivalperiodiccoherent}, the
revival peaks seem to be periodically spaced.

As a side note, there are models for which
Eq.~\eqref{atomicpopulationnewmodel} in fact describes the atomic
population dynamics
\cite{Buck:1981hi,Knight:1986kb,Phoenix:1990bk}. For example, Knight proposed a system where the levels $|g
\rangle$ and $|e \rangle$ are degenerate and connected by two-photon transitions through a
higher-energy virtual level \cite{Knight:1986kb}. The Hamiltonian
describing this process is
\begin{equation}
  \label{Hamiltonianalternative}
  \hat{H} = \omega \hat{a}^\dagger \hat{a} + g \hat{a}^\dagger \hat{a} \hat{\sigma}_x.
\end{equation}
With an initially excited atom, the inversion for this
model is precisely $W_p(t)$.

To set the scene for the next section, we introduce the
auxiliary function 
\begin{equation}
\label{revivalcomplex}
  Z(t) = \sum_{n = 0}^{+\infty} P_n e^{i 2 \pi f(n) t},
\end{equation}
which is the complex extension of both Eqs.~\eqref{atomicpopulation} and
\eqref{atomicpopulationnewmodel}. Setting $f(n)$ to $g \sqrt{n+1}/
\pi$ or $g n / \pi$ and taking the real part of $Z(t)$ yields $W(t)$
or $W_p(t)$, respectively. From now on, we will simply write $f(n)
\propto n$ or $f(n) \propto \sqrt{n+1}$. For simplicity, most of our computations
are done with $Z(t)$. 

While periodic revivals are very straightforward to understand,
quasi-periodic revivals have resisted a simple picture. Approximation schemes have been
developed for specific photon-number distributions \cite{Eberly:1980bw}, but the
incommensurable frequencies make Eq.~\eqref{atomicpopulation} quite
difficult to treat in general. On the other hand, the simplicity of periodic
revivals lies in that knowledge of a single revival cycle is
enough to generate the inversion
for any time.

The striking result we show in the next section is that a similar
picture actually holds true for the JCM quasi-periodic revivals: the
inversion is composed of underlying
packets. The packets are not perfect replicas of one
another, but knowing just one of them suffices to determines the
inversion completely. However, unlike the whole
inversion, the packets are usually localized in time, which makes them
more more useful for practical applications. We prove these claims in the
next section by introducing characteristic functions.

\section{Decomposing the inversion with the characteristic function}
\label{revivalcharacteristic}
In Sec.~\ref{sec:jcmmodel}, we briefly reviewed the JCM, its inversion quasi-periodic revivals, and some corresponding
periodic revivals. We introduced the auxiliary function $Z(t)$ in
Eq.~\eqref{revivalcomplex} to contemplate both kinds of revivals simultaneously. When $f(n) \propto n$ or $f(n) \propto \sqrt{n+1}$,
the real part of $Z(t)$ yields $W(t)$ or $W_p (t)$, respectively.
In this section,
we investigate $Z(t)$ further using characteristic functions.

The characteristic function of a probability distribution $P_n$ is
defined as the expectation value of $e^{i 2 \pi k n}$ \cite{Lukacs:108658}, \ie
\begin{equation}
  \label{characteristicfunction}
  \chi (k) \equiv \sum_{n=0}^{+\infty} P_n e^{i 2 \pi k n }.
\end{equation}
The function $\chi (k)$ is also simply a Fourier series with $P_n$ as coefficients.
It contains just as much physical information as $P_n$,
which is obtainable from $\chi (k)$ by inverting Eq.~\eqref{characteristicfunction}:
\begin{equation}
  \label{eq:45}
  P_n = \int_{-\frac{1}{2}}^{\frac{1}{2}} dk~ \chi (k) e^{-i 2\pi k n}.
\end{equation}
Due to $P_n$ being a discrete distribution, its characteristic
function, just like $W_p (t)$ in Eq.~\eqref{atomicpopulation}, is a periodic function: $\displaystyle\chi \left(k - \frac{1}{2} \right ) = \chi
\left( k + \frac{1}{2} \right)$.

It is not a mere coincidence that $\chi (k)$ and $W_p (t)$ are
both periodic: with the identification $\displaystyle k =
\frac{gt}{\pi}$, $W_p (t)$ is the real part of $\chi
(k)$. This follows from realizing that, when $f(n) \propto n$, the inversion complex
extension $Z(t)$ is the characteristic function itself: $Z (t) =
\chi ( gt / \pi)$.

When $f(n) \propto \sqrt{n+1}$, the connection between $\chi
(k)$ and $Z (t)$ is not so immediate. That said, we have seen in
Fig.~\ref{revivalcoherent} that the quasi-periodic revivals, to some
extent, have equally spaced peaks. This suggests that, even after replacing
commensurable frequencies by incommensurable ones, $Z (t)$ seems to
still inherit some properties of $\chi (k)$, such as its periodicity, to a certain degree. Our goal is to put this connection on
more precise grounds by expressing $Z(t)$ in terms of $\chi(k)$ for a general $ f(n)$.

With this goal in mind, we introduce the distribution
  \begin{equation}
\label{continuousdistribution}
  P(x) \equiv \sum_{n=0}^{+\infty} P_n \delta (x - n )
\end{equation}
which allows us to rewrite 
rewrite $Z(t)$ as
\begin{equation}
\label{revivalasft}
  Z(t) = \int_{-\infty}^{+\infty} dx P(x) e^{i 2\pi f(x) t },
\end{equation}
where $f(x)$ is just the extension of $f(n)$ to real numbers, \eg~ $n \to
x$ and $\sqrt{n+1} \to \sqrt{x+1}$. It is easy to check from
Eq.~\eqref{characteristicfunction} that the distribution $P(x)$ also
has $\chi (k)$ as its characteristic function:
\begin{equation}
  \label{eq:69}
  P(x) = \int_{-\infty}^{+\infty} dk~\chi (k) e^{-i 2\pi k x}.
\end{equation}
Unlike Eq.~\eqref{eq:45}, the integral in
Eq.~\eqref{eq:69} is not bounded, which will be useful in the next
steps.

We now substitute from Eq.~\eqref{eq:69} into Eq.~\eqref{revivalasft} to express
$Z(t)$ in terms of $\chi (k)$:
\begin{equation}
\label{integralZZ}
  Z(t) = \int_{-\infty}^{+\infty} dk~\chi(k) \int_{-\infty}^{+\infty}
  dx~ e^{i 2 \pi \left[ f (x) t - k x \right] }.
\end{equation}
The integral over $x$ is some distribution dependent on $k$ and
$t$, which we will denote $\mathcal{K} (k, t)$:
\begin{equation}
  \label{definitionKernel}
  \mathcal{K} (k, t) \equiv \int_{-\infty}^{+\infty}
  dx~ e^{i 2 \pi \left[ f (x) t - k x \right] }.
\end{equation}
It may be interpreted as a propagator that determines $Z(t)$, given
$\chi (k)$. Unlike $\chi (k)$, the propagator is not necessarily periodic with
respect to $k$. This ultimately leads to $Z(t)$ not being exactly
periodic in general.

Next, we use the periodicity of $\chi (k)$ to split the integral over $k$ in Eq.~\eqref{integralZZ} into
a sum of integrals, each of which ranging from $ m-\frac{1}{2}$ to $ m+ \frac{1}{2}$, with $m \in \mathbb{Z}$. Then, for each
interval, we make the change of variables $k \to
k+m$, so that each integral covers the same range $[-1/2, 1/2)$.
Eq.~\eqref{integralZZ} then simplifies to
\begin{align}
  \label{eq:44}
    Z(t) &= \sum_{m=-\infty}^{+\infty} Z_m (t), \\
\label{Zmm} Z_m (t) &= \int_{-\frac{1}{2}}^{\frac{1}{2}}  dk~ \chi (k) \mathcal{K} (k
    +m , t),
\end{align}
where the periodicity of $\chi(k)$ allows us to replace $\chi(k+m)$
by $\chi(k)$.

The decomposition in Eq.~\eqref{eq:44} involves no approximation. We now specialize Eq.~\eqref{Zmm} for the cases
$f(n) \propto n$ and $f(n) \propto \sqrt{n+1}$. We show that, in
both cases, knowledge of a single $Z_m (t)$ is enough to determine $Z
(t)$. 

\subsection{The case $f(n) \propto n$}
When $f(n) \propto n$, it is easy to verify that $\mathcal{K} (k+m, t)$ is simply $\displaystyle \delta\left( k+m - \frac{gt}{ \pi}\right)$. Since $k$
only ranges from $-1/2$ to $1/2$ in Eq.~\eqref{Zmm}, $gt/\pi$ must be
in the range of $m -1/2$ and $m+1/2$ for $\displaystyle \delta\left( k+m - \frac{gt}{\pi}\right)$ to
contribute. Therefore, for a given $t$, only a single $Z_m (t)$ is not zero, and happens to be the
characteristic function when we use the delta function to integrate: 
\begin{equation}
\label{zMPERiodic}
Z_m (t) =  \chi\left( \frac{gt}{\pi} \right) \Pi \left( \frac{gt}{\pi} - m \right),
\end{equation}
where $\Pi (x)$ is the rectangular function, equal to $1$ for $-1/2 <
x < 1/2$ and $0$ otherwise.

Hence, each $Z_m (t)$ is a replica of every other one, centered at $g t_m = m
\pi$. In Fig.~\ref{revivalperiodiccoherent}, each revival corresponds to the real part of a different $Z_m (t)$. We show next that, for $f(n) \propto
\sqrt{n+1}$, a similar picture also holds true, except that the $Z_m
(t)$ are not perfect copies of one another: they also experience dispersion akin to that of quantum-mechanical
wave packets. 

\subsection{The case $f(n) \propto \sqrt{n+1}$}
When $f(n) \propto \sqrt{n+1}$, the propagator $\mathcal{K} (k, t)$ is more complicated, but the decomposition of $Z (t)$
as a sum of $Z_m (t)$ remains exact. The non-periodicity of
$\mathcal{K}(k, t)$ implies that $Z_m (t)$ now actually depends on $m$. Also, since $\mathcal{K} (k, t)$ is not a delta
function, the simple identification $k = gt / \pi$ found for the case
$f(n) \propto n$ does not hold. 

In spite of such complications, as we continuously 
deform $f(n)$ from $n$ to $\sqrt{n+1}$, we expect periodic revivals
such as the ones in
Fig.~\ref{revivalperiodiccoherent} to gradually yield place
to the quasi-periodic revivals such as the ones in Fig.~\ref{revivalcoherent}. If these revivals do not
overlap during the process, it is natural to associate each
JCM revival to a single $Z_m (t)$. In this scenario, the first collapse, in
particular, would be identified with
\begin{equation}
  \label{collapsestructure}
  Z_0 (t) = \int_{-1/2}^{1/2} dk~ \chi (k) \int_{-\infty}^{\infty}
  dx~e^{i 2\pi \left[ f(x) t - k x \right]}.
\end{equation}

A technical detail worth mentioning is that the definition of $Z_0 (t)$ must
encompass a portion of the region $t < 0$. This is easier to justify through periodic revivals. In Fig.~\ref{revivalperiodiccoherent}, we see that the region near $t = 0$ only
comprehends half of the structure replicated at later
times, so we must extend $Z(t)$ for $t < 0$ to capture
the missing half. This must, then, also be true for quasi-periodic
revivals. On the other hand, in an experimental setup, one may
measure only $W(t) = \Re \left\{ Z(t) \right\}$, and only for $ t >
0$. However, since $W(t)$ is an even function, it can be
readily extended to $t < 0$.

We now show that a single $Z_m (t)$ has complete information about the whole
$Z(t)$. This is more easily seen in Fourier space, where it will be
clear that the Fourier transform of each $Z_m (t)$ differs only by a
phase from every other one. We define the Fourier transform of $Z_m (t)$
as
\begin{equation}
  \tilde{Z}_m (\nu) \equiv \int_{-\infty}^{+\infty} dt~ e^{i 2\pi \nu t} Z_m (t). 
\end{equation}
Then, Fourier-transforming both sides of Eq.~\eqref{Zmm}, it follows that
\begin{equation}
  \label{fouriertransformrevival}
  \tilde{Z}_m (\nu) = \int_{-1/2}^{1/2} dk~\chi (k) \int_{-\infty}^{\infty}
  dx~ \delta\left[f(x) - \nu \right] e^{-i 2\pi  (k+m) x },
\end{equation}

The integral over $x$ can be readily performed by using the
property $\displaystyle \delta \left[ g(x)\right] = \sum_i \frac{\delta (x -
    x_i)}{g'(x_i) }$, where $x_i$ are the roots of $g(x)$. For $f(x) = g \sqrt{x+1}/\pi$, we
  have a single root, $x = \displaystyle\left(\frac{\pi
      \nu}{g}\right)^2 - 1$, and only if $\nu > 0$. If $\nu < 0$, there is no
  solution, which implies that $\tilde{Z}_m (\nu) = 0$ for $\nu <
  0$. This also follows from $Z(t)$ being a sum of only positive
  frequencies signals, according to Eq.~\eqref{revivalcomplex}.

Finally, after integrating the right-hand side of
Eq.~\eqref{fouriertransformrevival} over $x$ and moving every factor
independent of $k$ to the left-hand side, we get
\begin{align}
\left(\frac{g}{2 \pi  } \right)^2 &\frac{\tilde{Z}_m (\nu)}{\nu} e^{i 2 \pi m \left(\frac{ \pi
        \nu}{g} \right)^2} = \nonumber \\
\label{Znualreadyintegratedx} & \int_{-1/2}^{1/2} dk~\chi (k) e^{-i 2\pi
    k \left[ \left(\frac{\pi \nu}{g} \right)^2 - 1 \right]},
                                \quad\text{if }\nu > 0; \nonumber \\
&  0, \quad \text{otherwise}. 
\end{align}
Intriguingly, only the left-hand side of
Eq.~\eqref{Znualreadyintegratedx} depends on $m$. This means that
each $\tilde{Z}_m(\nu)$ can differ only by a phase from one another:
\begin{equation}
  \label{relationbetweenrevivals}
  \tilde{Z}_m (\nu) = \tilde{Z}_0 (\nu) e^{-i 2\pi m\left(\frac{\pi \nu}{g}\right)^2}.
\end{equation}
This relation is one of the major results of this work, and we now
discuss its implications. The phase we just encountered depends on $\nu$ quadratically. If the dependence
were linear, this phase would simply translate $Z_0
(t)$ in time. However, it is well-known from quantum dynamics of
free particles that quadratic dependencies lead
to an overall translation, but also to some dispersion. 

To make this analogy clearer, consider a quantum-mechanical wave packet
in free space $\psi
(x)$, and its Fourier transform $\tilde{\psi} (p)$, describing a
particle of mass $\mu = 1/2$. The time-evolved
$\psi (x, \uptau)$ is obtained by taking the inverse Fourier transform
of $\displaystyle \tilde{\psi} (p) e^{ -i \uptau p^2 
}$. The analogy goes as follows:
$x$, $\psi(x)$, $p$ and $\tilde{\psi} (p)$  correspond to $t$, $Z_0
(t)$, $\nu$ and $ \tilde{Z}_0 (\nu)$, respectively; and $Z_m (t)$
corresponds to $\psi (x, \uptau)$, with $m$ determining the effective
$\uptau$. 

In quantum mechanics, a packet propagates over space and disperses as it
moves. Each $Z_m (t)$ is analogous to a snapshot of
the wave packet. The inversion is the superposition of the
snapshots. Nonetheless, a single snapshot $Z_m (t)$ is enough to determine
every other snapshot, just as knowledge of the
quantum-mechanical wave packet for some instant implies knowledge of it for any other
instant. Hence, a single $Z_m (t)$ determines the inversion completely.

Therefore, while $Z(t)$ and $W(t)$ may, in general, look very
irregular and complicated, it should be possible to distill them and
identify an underlying pattern corresponding to the juxtaposition of
different $Z_m (t)$ or $W_m (t) \equiv \Re\left\{ Z_m (t)\right\}$. Next, we illustrate the distillation for a coherent
state with average photon-number of $20$.
 The first step is to
identify $Z_0 (t)$. We already argued that, if the collapse and
the first revival do not overlap, we may
associate $Z_0 (t)$ with the collapse. We then numerically compute $\tilde{Z}_0
(\nu)$ through the Fast Fourier Transform (FFT) method
\cite{Press:1992uq}. With Eq.~\eqref{relationbetweenrevivals}, we find the subsequent $\tilde{Z}_m (\nu)$. Finally, the inverse FFT of $\tilde{Z}_m (\nu)$
yields $Z_m (t)$.

\begin{figure}[h]
\includegraphics[width=1\columnwidth]{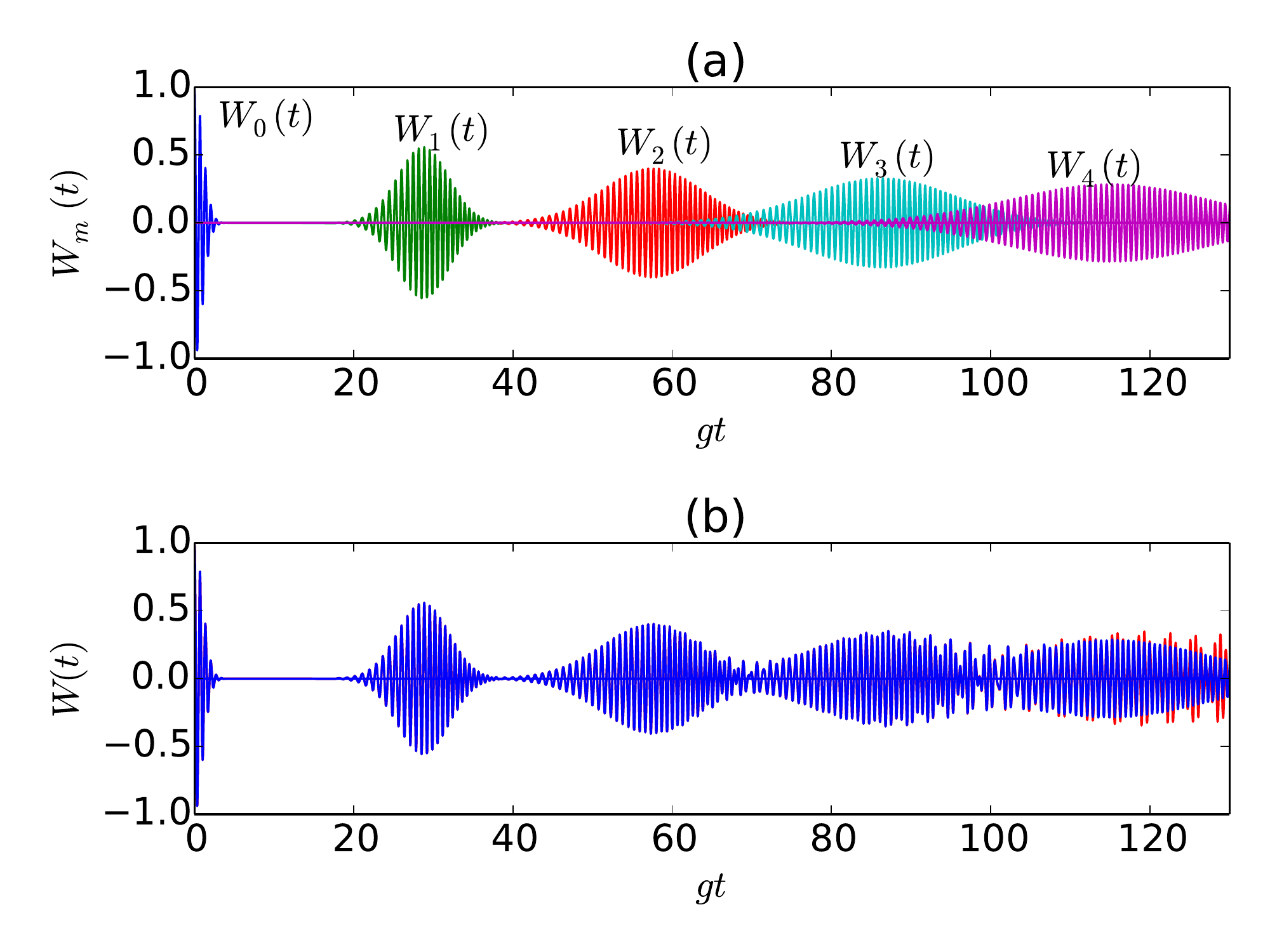}
\caption[distillcoherent]{\label{distillcoherent} We consider the
  revivals for a coherent state with average photon-number of $20$ in
  light of the decomposition of Eq.~\eqref{eq:44}. In
  Fig.~\ref{distillcoherent} (a), we juxtapose different $W_m
  (t)$. According to Eq.~\eqref{relationbetweenrevivals}, they become
  the same once dispersion is accounted for. The kind of
  dispersion is the same experienced by free wave packets in quantum
  mechanics. In Fig.~\ref{distillcoherent} (a), we compare the result
  of adding these replicas (blue curve) and the actual population
inversion (red curve). There is excellent agreement, except for longer
times ($gt \sim 110$), which may be fixed by adding subsequent $W_m (t)$.}
\end{figure}

We present the results of this procedure in
Fig.~\ref{distillcoherent}, where we have considered $m = 0, 1, 2, 3,
4$. In Fig.~\ref{distillcoherent} (a), we
simply juxtapose (the real part of) each $Z_m (t)$. In Fig.~\ref{distillcoherent} (b), we add them
up. Here, the blue
curve corresponds to $\displaystyle \sum_{m=0}^4 W_m (t)$ and the
red curve (visible only after $gt \sim 110$) is the actual $W
(t)$, calculated numerically with Eq.~\eqref{atomicpopulation}. The
agreement can be improved for longer times by adding more $W_m (t)$.

Following the analogy with quantum mechanics, we estimate at what
time each $Z_m (t)$ is centered. Whereas in quantum mechanics one
linearizes the dispersion relation, here we linearize the phase in
Eq.~\eqref{relationbetweenrevivals} around some frequency $\tilde{\nu}$ at which $\tilde{Z}_0 (\nu)$ is
peaked. With $Z_0 (t)$ centered at
$0$, $Z_m (t)$ should be centered at
\begin{equation}
  \label{eq:2}
   t_m = 2 m\left(\frac{\pi }{g}\right)^2 \tilde{\nu}.
\end{equation}
There must correspond a photon
number to the dominant frequency $\tilde{\nu}$, which we define through $2 \pi \tilde{\nu} \equiv 2g
\sqrt{\tilde{n}+1}$. In terms of $\tilde{n}$,
\begin{equation}
  \label{eq:3}
 t_m = \frac{2 \pi \sqrt{\tilde{n}+1}}{g} m,
\end{equation} 
Naturally, these times also correspond to when the terms of
Eq.~\eqref{atomicpopulation} are approximately in phase \cite{Yoo:1985uq}.
In a loose sense, $\tau \equiv \displaystyle \frac{2 \pi \sqrt{\tilde{n}+1}}{g}$
works as a period, except that $Z_m (t)$ also widens as we increase
$m$. This explains why the revival peaks in Fig.~\ref{revivalcoherent}
are approximately equally spaced. Since the $Z_m (t)$ are ultimately are generated by $\chi (k)$, it is natural
to make the more general identification
\begin{equation}
  \label{timeandkjcm}
 t  \to \frac{2 \pi \sqrt{\tilde{n}+1}}{g} k,
\end{equation} 
so that, when we increment $k$ by one, $t$ also changes by $\tau$.

To finish this section, we discuss how to obtain the
probabilities from a single $Z_m (t)$, a relevant task in the context of quantum
tomography. Firstly, we notice that the right-hand sides of
Eqs.~\eqref{Znualreadyintegratedx} and \eqref{eq:45} have very similar
forms. It follows that, by setting $\nu = \nu_n$, where
$2 \pi \nu_n \equiv 2g \sqrt{n+1}$, the right-hand side of
Eq.~\eqref{Znualreadyintegratedx} becomes simply $P_n$. On the left-hand side of
Eq.~\eqref{Znualreadyintegratedx}, replacing $\nu$ by $\nu_n$
eliminates the phase factor. We are then left simply with
\begin{equation}
  \label{probabilityfinally}
  P_n = \frac{g^2}{2 \pi^2} \frac{\tilde{Z}_m (\nu_n)}{\nu_n}.
\end{equation}
Notice that $2 \pi \nu_n = 2g \sqrt{n+1}$ are the oscillation
frequencies of the oscillators in Eq.~\eqref{atomicpopulation}.

It is interesting to rewrite Eq.~\eqref{probabilityfinally} in terms
of the actually measurable $W(t)$. Firstly, we define $W_m (t) =
\Re\left\{ Z_m(t) \right\}$, and $\tilde{W}_m
(\nu)$ as its Fourier transform. Then it is not hard to
show that $\tilde{W}_m
(\nu) = \displaystyle\frac{ \tilde{Z}_m (\nu)}{2}$ for $\nu > 0$. Therefore,
\begin{equation}
  \label{probabilityfinallyreal}
  P_n = \frac{g^2}{\pi^2} \frac{\tilde{W}_m (\nu_n)}{\nu_n}.
\end{equation}
A similar relation was previously obtained through a Poisson sum formula
approach under the assumption of non-overlapping revivals \cite{Fleischhauer:1993dw}. Eq.~\eqref{probabilityfinallyreal} states
that the probability distribution is codified in the frequencies of
$W_m (t)$
present in Eq.~\eqref{atomicpopulation}.

Nevertheless, the relation between
probabilities and $W_m(t)$ is exact, since it follows from
Eqs.~\eqref{eq:44} and \eqref{Zmm}. 
However, experimentally, only the whole
$W(t)$ can be measured. Thus, having non-overlapping revivals is more
of a convenience, as it allows us to approximately identify the $m$-th revival of $W
(t)$ to $W_m (t)$ immediately. In this regime, our method is equivalent to
that Ref.\cite{Fleischhauer:1993dw}. However, particularly for low average
photon-numbers, this approximation breaks down already for $m = 0$. It
is then that our formalism shines: we use it in Sec.~\ref{overlapping}
to cast away this limitation and retrieve $W_0
(t)$ even when revivals overlap.

\begin{figure}[h]
\includegraphics[width=1.\columnwidth]{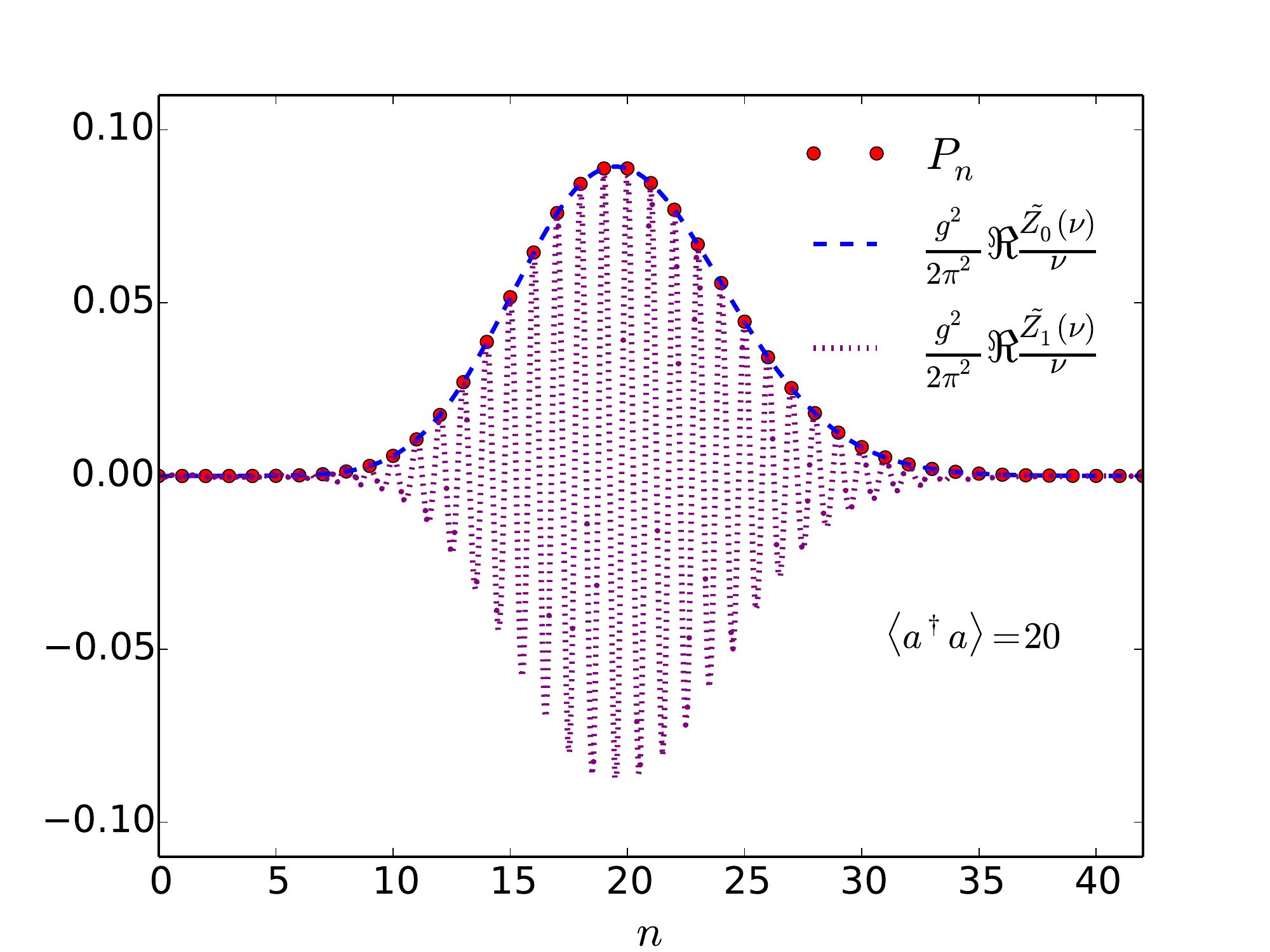}
\caption[revivaltestcoherent]{\label{revivaltestcoherent} Comparison
  between the photon-number distribution $P_n$ of a coherent
  state with average photon-number of $20$ and the functions $\displaystyle \frac{g^2}{\pi^2} \frac{\tilde{W}_m
  (\nu)}{\nu} = \frac{g^2}{2\pi^2}  \frac{\Re \{\tilde{Z}_m
  (\nu) \}}{\nu}, m=0,1$, showing the validity of
Eq.~\eqref{probabilityfinallyreal}. It is also worth mentioning that the
function with $m = 1$ is enveloped by the one with $m = 0$, in
agreement with Eq.~\eqref{relationbetweenrevivals}.}
\end{figure}

We now use $W_0 (t)$ and $W_1 (t)$, previously shown in
Fig.~\ref{distillcoherent}, to retrieve the photon-number distribution
of a coherent state with $\langle a^\dagger a \rangle = 20$. The results
are shown in Fig.~\ref{revivaltestcoherent}. The red circle-shaped dots are the theoretical $P_n$. The dashed curves are
the right-hand side of
Eqs.~\eqref{probabilityfinallyreal} for $m = 0$ and $m = 1$. We
convert the argument $\nu$ to $n$ through the identification $2\pi \nu = 2g \sqrt {n+1}$. Eq.~\eqref{probabilityfinallyreal} predicts that,
when $n$ is an integer, the plotted function should match $P_n$, which
is consistent with the behaviour of the dashed curves. Moreover, the
faster-oscillating purple line, which corresponds to $m = 1$, has
the blue line ($m = 0$) as its envelope, in agreement with
Eq.~\eqref{relationbetweenrevivals}. 

\section{Distilling the revivals of a cat state}
\label{understandingcats}
In the previous section, we have shown that the JCM population revivals are a
result of interference between a set of packets $Z_m (t)$, which
are akin to snapshots of a quantum-mechanical wave packet for different
times. We illustrated this decomposition explicitly for a coherent
state. Its distribution being relatively steady, the characteristic
function of a coherent state is peaked around $k \sim m$, $m \in \mathbb{Z}$. For
periodic revivals, where $Z(t) = \chi (gt/\pi)$, this translates to $W(t)$ being peaked around $g t
= m \pi$, as shown previously in
Fig.~\ref{revivalperiodiccoherent}. For quasi-periodic revivals, $Z(t)$
and $\chi (k)$ are not directly proportional, but the propagator
$\mathcal{K} (k, t)$ defined in Eq.~\eqref{definitionKernel} maps $\chi
(k)$ within one of its periods into one of the $W_m (t)$ shown in Fig.~\ref{distillcoherent}
(a). They are centered at the linearly-spaced intervals dictated by
Eq.~\eqref{eq:3}.

We wish to illustrate how to decompose other inversion profiles, and a cat state is a natural choice, given its
relative simplicity. A cat state is usually defined as
\begin{equation}
  \label{catstatedefinition}
  |\xi \rangle \propto \frac{| \alpha \rangle + |\alpha e^{i \phi} \rangle}{\sqrt{2}},
\end{equation}
where  $| \alpha \rangle$ is a coherent state with $\langle a^\dagger a
\rangle = | \alpha |^2$. The symbol $\propto$ indicates
that we have not normalized the state properly, though the
missing proportionality factor approaches $1$ for large $\alpha$. We assume for simplicity
that $\alpha$ is real.

The photon-number distribution of $| \xi \rangle$ is 
\begin{equation}
  \label{Pncatstate}
  P_n (\phi ) \approx  c^2_n  + c^2_n \cos \left(n\phi\right) ,
\end{equation}
where $c^2_n$ is the photon-number distribution of a coherent
state, and we have assumed large $\alpha$. The second term on the
right-hand side of Eq.~\eqref{Pncatstate} oscillates with frequency dictated by
$\phi$. The oscillations are fastest when $\phi = \pi$, in which case $P_n$
alternates between $2 c^2_n$ (for even $n$) and 0 (for odd $n$). The
characteristic function, being essentially a spectral decomposition of
$P_n$, should be peaked around $k \sim 0$ (accounting for the steadier
component of $\chi (k)$) and around $\displaystyle k \sim \frac{\phi}{2\pi}$ (accounting for the
staggered component of $\chi (k)$). In fact, let $\chi_{\alpha} (k)$ be the
characteristic function for $| \alpha \rangle$. Then, from
Eq.~\eqref{Pncatstate}, the
characteristic function for $| \xi \rangle$ is
\begin{equation}
  \label{characteristcfunction}
  \chi (k, \phi) = \chi_\alpha (k) + \frac{1}{2} \chi_\alpha \left( k -
    \frac{\phi}{2\pi} \right) + \frac{1}{2} \chi_\alpha \left( k +
    \frac{\phi}{2\pi} \right).
\end{equation}
In particular, for $\phi = \pi$, and already using the periodicity
of $\chi (k, \phi)$,
\begin{equation}
  \label{characteristicagain}
  \chi (k, \pi) = \chi_\alpha (k) +  \chi_\alpha \left( k -
    \frac{1}{2} \right).
\end{equation}
Due to the second term on the right-hand side of Eq.~\eqref{characteristicagain}, this characteristic function features
additional peaks around $k \sim m+ 1/2$, $m \in \mathbb{Z}$.

\begin{figure}[h]
\includegraphics[width=1.\columnwidth]{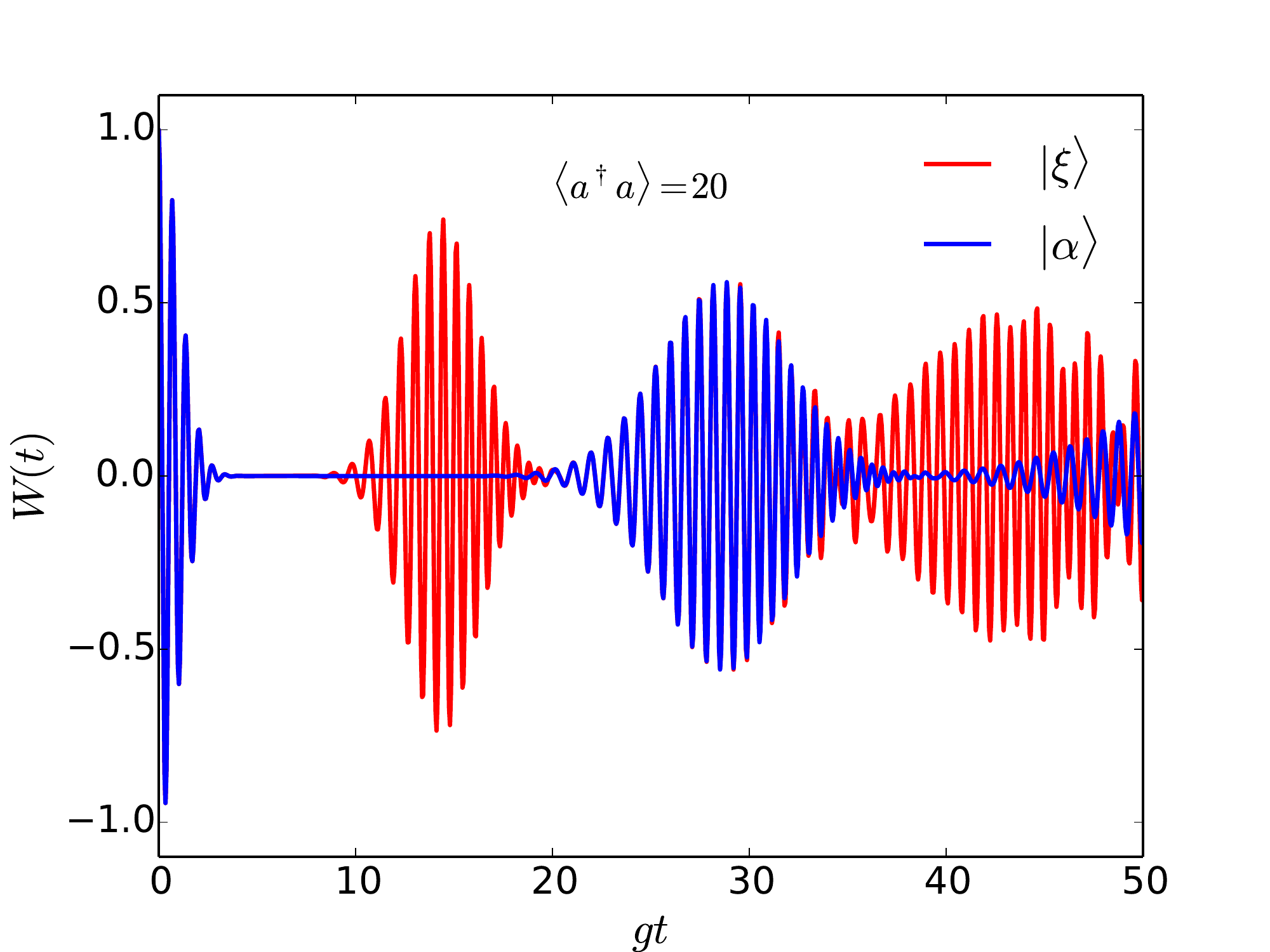}
\caption[revivalcat]{\label{revivalcat} Revivals of
a coherent state $|\alpha\rangle$ and a cat state $| \xi \rangle \propto
| \alpha \rangle + | -\alpha \rangle$, with $\langle a^\dagger a
\rangle = |\alpha|^2 = 20$. Since the characteristic function of these
states is the same for low-frequencies, the collapse and some revivals
coincide. However, the high-frequency components exclusive to $| \xi
\rangle$ generate intermediate revivals arising from the oscillatory behaviour of the photon-number
distribution of a cat state.}
\end{figure}

Let us now look at the revivals of $| \xi \rangle$ and how they
compare to revivals of $| \alpha \rangle$ in Fig.~\ref{revivalcat}. Their initial
collapses turn out to be, to a very good approximation, the same. At first, this is very unsettling: if
we take the collapse as data for $W_0 (t)$, a naive application of Eq.~\eqref{probabilityfinallyreal} will then
yield (incorrectly) the distribution of a coherent
state. On the other hand, the revivals of $| \xi \rangle$ seem to
happen earlier. One may numerically check that using the first
revival as data for $W_1 (t)$ leads to unphysical (negative) probabilities.

These puzzling features can be understood almost immediately in the framework
of characteristic functions. To shed light on the matter, we first
analyze periodic revivals,
then argue that, as we deform $n \to \sqrt{n+1}$, the quasi-periodic
revivals must remain qualitatively similar. When revivals are periodic, they
are proportional to $\chi (k)$ itself. On the other hand, we expect
$\chi(k)$, according to Eq.~\eqref{characteristicagain}, to be peaked around $m$
and $m - 1/2$, with $m$ an integer. Within a single period window, this corresponds to
two peaks. As we now look at the JCM revivals, it is natural to still expect
two peaks, albeit with modified shapes. This means that we should
interpret the first collapse and the first revival of $| \xi \rangle$
seen in Fig.~\ref{revivalcat} as associated to a single period of
$\chi (k)$ rather than separate objects.

This picture also explains why the initial collapses of both cat and
coherent states overlap: the initial collapses being associated to the
peak of $\chi (k)$ centered at $k \sim 0$, they are the same for
states $|\alpha\rangle$ and $| \xi \rangle$ because the steadier components
of the characteristic functions of both states are the same. On the other hand, the
extra revivals of the cat state corresponds to the staggered components
of $\chi(k, \pi)$, which are absent for a coherent state.  

\begin{figure}[h]
\includegraphics[width=1.\columnwidth]{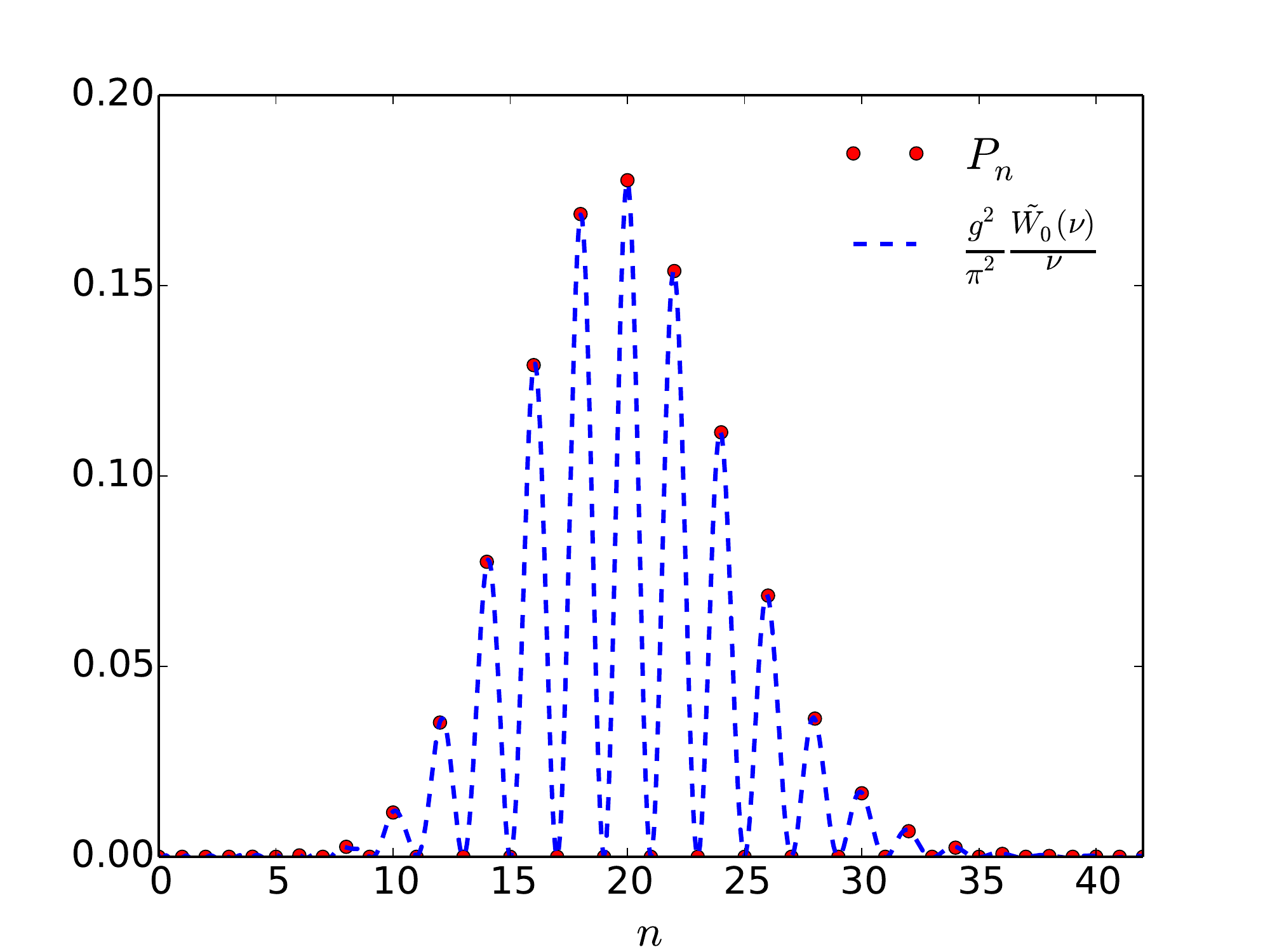}
\caption[revivalcat]{\label{probcat} Comparison between the
  photon-number distribution of a cat state and
  Eq.~\eqref{probabilityfinallyreal} for $m = 0$. When collapse (and
  its symmetric extension for $t < 0$) and the first revival are accounted
  as $W_0 (t)$, one retrieves the correct distribution.}
\end{figure}

\begin{figure}[h]
\includegraphics[width=1.\columnwidth]{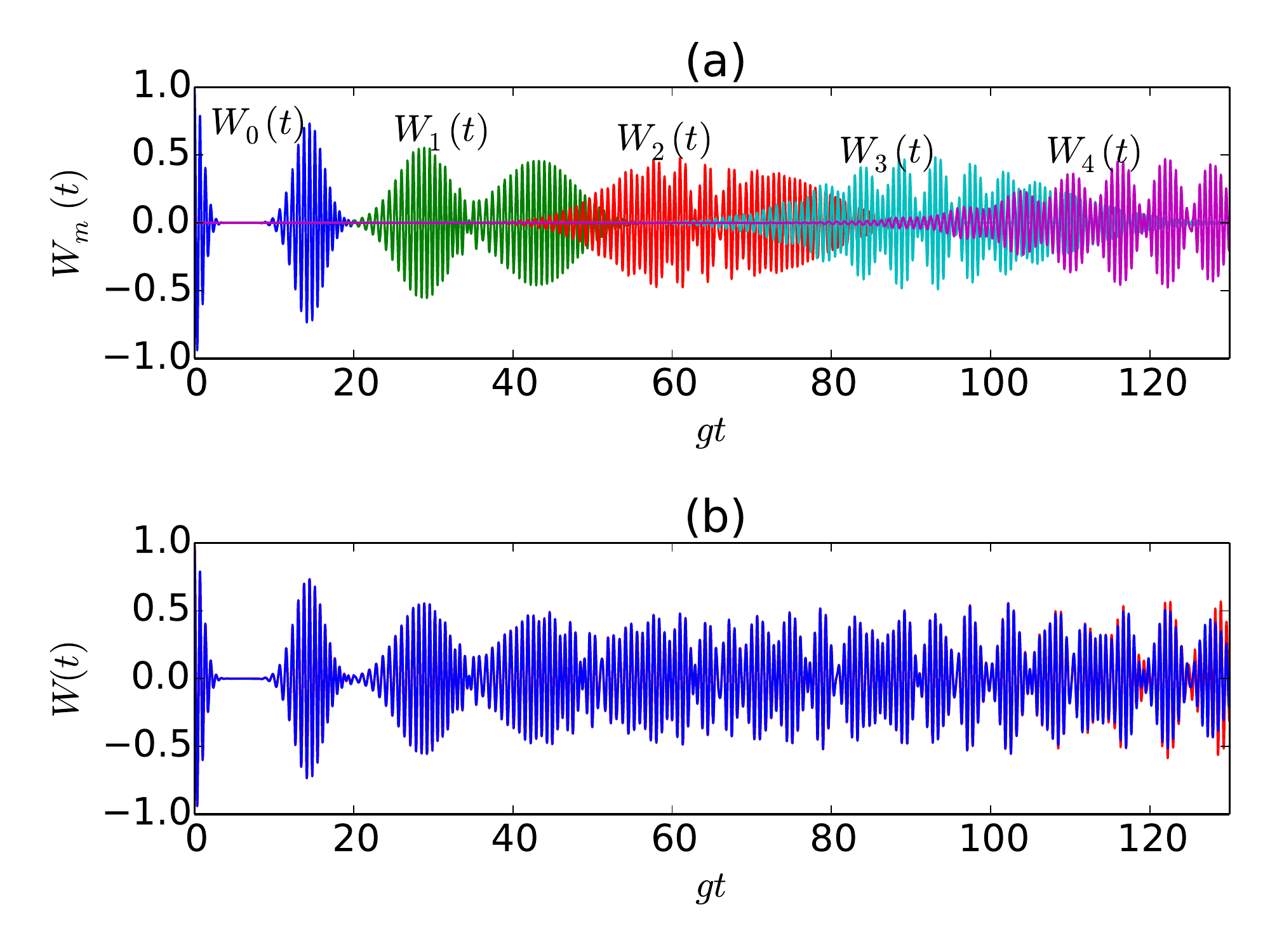}
\caption[revivalcat]{\label{distillcat} (a) Decomposition of the
  revivals of $| \xi \rangle$ as given by
  Eq.~\eqref{catstatedefinition}, with $\phi = \pi$, in terms of $W_m
  (t)$. Two consecutive revivals are part of the same $W_m (t)$. In the
  framework of characteristic functions, each correspond to the peaks
  of $\chi(k)$ centered around $m$ and $m + 1/2$. (b) When we add the
  $W_m (t)$, we recover the full $W(t)$. The blue curve represents
  $\sum_{m=0}^4 W_m (t)$ and the red curve represents the full
  $W(t)$. The agreement can be improved for even longer times by adding
  further $W_m (t)$.}
\end{figure}

In conclusion, both the collapse and this early revival
must be interpreted as $W_0 (t)$. It is only when they are taken
simultaneously into consideration that
Eq.~\eqref{probabilityfinallyreal} yields the correct probabilities,
as shown in Fig.~\ref{probcat}. In Fig.~\ref{distillcat} (a), we show
the $W_0 (t)$ used to retrieve the probabilities and the other $W_m (t)$
generated through Eq.~\eqref{relationbetweenrevivals}. In
Fig.~\ref{distillcat} (b), we add them to show that we recover the full $W(t)$.

A practical task is how to verify when revival is
not a standalone $W_m (t)$. In our example we could
simply compare the predicted probabilities, but we do not
know the state behind the inversion profile in general. One approach
to tackle this task is simply through trial and error: once a
distribution has been predicted, $W(t)$ can be numerically calculated with
Eq.~\eqref{atomicpopulation} and compared with the experimental data
for the inversion at later times. If comparison shows that the
numerically calculated inversion is missing intermediate revivals, then the time range for $W_m (t)$ must be reselected, just as we did for a cat
state.  

Another check is that, since the additional
revivals are not standalone $W_m (t)$, using just the additional revival in Eq.~\eqref{probabilityfinallyreal}
will, in general, yield negative, hence non-physical,
probabilities. They only make sense when added to the
ones obtained through the collapse. Their combination should yield the
correct probabilities.

\section{Overlapping $W_m (t)$}
\label{overlapping}
In the previous sections, we considered non-overlapping revivals so that $W_0 (t)$ (or any other $W_m (t)$) can be obtained straightforwardly from $W (t)$, and Eq.~\eqref{probabilityfinallyreal} used
to retrieve the field photon-number distribution. We now show that, with the decomposition presented in Sec.~\ref{revivalcharacteristic}, it is possible to circumvent this limitation and extract $W_0
(t)$ even when revivals overlap. Therefore, our method allows
photon-number distributions to be retrieved under fairly general
conditions.

The key idea is that, if we sample $W(t)$ for a long enough time
window $[-T, T]$, $W_0 (t)$ will be completely captured, albeit
tainted with tails of $W_1 (t)$, $W_{-1} (t)$ and, more generally, every
other $W_m (t)$. Here, we consider the simplest case where terms with
$| m | > 1$ can be neglected, though the formalism is easy to
acommodate otherwise. Considering the FFT of the limited window of $W(t)$
leads to the spectrum $\tilde{\mathcal{W}}(\nu)$:
\begin{align}
\label{integralequation}
  \tilde{\mathcal{W}} (\nu) &=~ \tilde{W}_0 (\nu) \\
&+ \int_{-\infty}^{\infty} d\nu' ~ \frac{\sin\left[ \pi (\nu - \nu') 2T \right]}{\pi (\nu -
  \nu')} \left[ \tilde{W}_1 (\nu') +
\tilde{W}_{-1} (\nu') \right], \nonumber
\end{align}
The first term on the right-hand side of Eq.~\eqref{integralequation}
is simply the spectrum of $W_0 (t)$. However, $\tilde{\mathcal{W}} (\nu)$ is contamined by the
second term, which arises from the convolution of $W_1 (t)$
and $W_{-1} (t)$ with the window function located between $-T$
and $T$. 

\begin{figure}[t]
\includegraphics[width=1.\columnwidth]{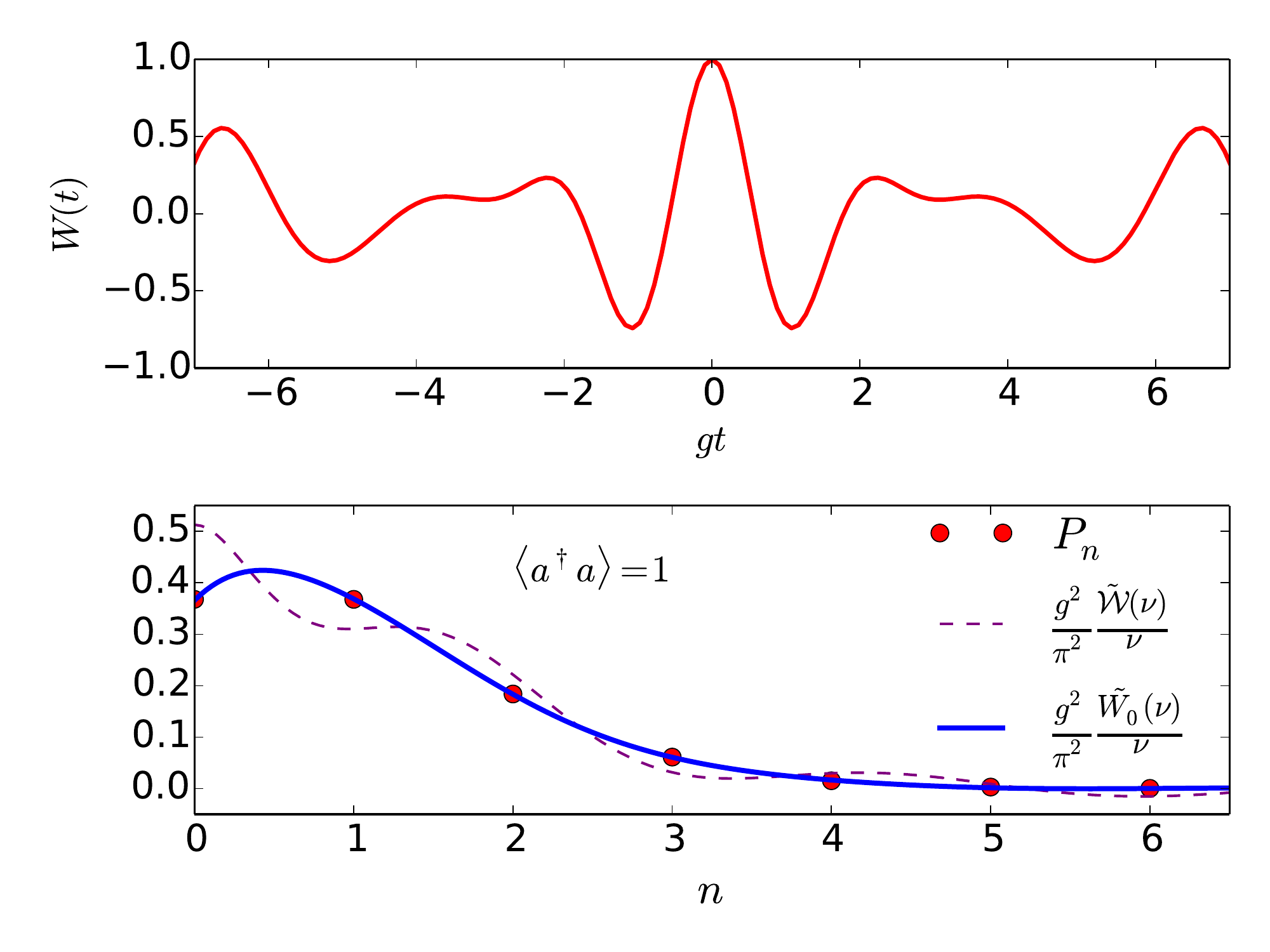}
\caption[revivaln1]{\label{revivaln1} 
Inversion profile for a coherent state with $\langle a^\dagger
a \rangle = 1$ and photon-number distribution retrieval through
Eq.~\eqref{integralequationfinal}. For an average photon-number this
low, the revivals cannot be clearly resolved anymore. We tentatively
choose a time window running from $-5$ to $5$. The Fourier transform
of $W(t)$ under this window yields $\tilde{\mathcal{W}} (\nu)$, which
we use to construct the (dashed) purple curve. This curve does not
match the theoretical probability (represented by the red dots)
because it is contaminated by the spectra of $W_1 (t)$ and $W_{-1}
(t)$. We have used Eq.~\eqref{integralequationfinal} to solve for $\tilde{W}_0
(\nu)$, which corresponds to the blue curve, yielding the correct probabilities.}
\end{figure}

However, from Eq.~\eqref{relationbetweenrevivals}, it may be shown that $\displaystyle \tilde{W}_1 (\nu) +
\tilde{W}_{-1} (\nu) = 2 \tilde{W}_0 (\nu) \cos \left[ 2 \pi\left(\frac{\pi \nu}{g} \right)
  \right]^2 \tilde{W}_0 (\nu)$, which means that
  Eq.~\eqref{integralequation} is, in fact, an integral equation for
  $\tilde{W}_0 (\nu)$:
\begin{align}
  \label{integralequationfinal}
  \tilde{\mathcal{W}} (\nu) = \tilde{W}_0 (\nu) + \int_{-\infty}^{\infty} d\nu' ~ S (\nu, \nu') \tilde{W}_0 (\nu'), 
\end{align}
where
\begin{equation}
  S(\nu, \nu') = 2 \cos \left[ 2 \pi\left(\frac{\pi \nu'}{g} \right)
  \right]^2 \frac{\sin\left[ \pi (\nu - \nu') 2T \right]}{\pi (\nu -
  \nu')}.
\end{equation}
This equation has the form of a Fredholm equation of the second kind,
and can be solved numerically  for $\tilde{W}_0 (\nu)$, given the
observed spectrum $\tilde{\mathcal{W}} (\nu)$ \cite{Press:1992uq}. We consider the retrieval of $\tilde{W}_0 (\nu)$, and the photon-number
distribution through Eq.~\eqref{probabilityfinallyreal}, for a
coherent state with $n = 1$, for which revivals cannot be
resolved. The results are presented in
Fig.~\ref{revivaln1}. Eq.~\eqref{integralequationfinal} lays the
foundation for our approach. A more detailed exposition of this
technique will be presented in a later work.

\section{Conclusions}
\label{sec:summary}
In this work, we exploited
the characteristic function periodicity to split the inversion into a
superposition of packets centered at different times. When the inversion oscillation frequencies are
commensurable, the packets are perfect replicas of one another, and
each one represents a single revival. 
In the case of the JCM, for which frequencies are incommensurable,
the inversion $W(t)$ can still be split exactly into a set of packets $W_m
(t)$. Knowledge of a single $W_m (t)$ determines every other $W_m
(t)$, but they are now imperfect replicas, experiencing
dispersion akin to that of free particles in quantum dynamics. Once
dispersion is accounted for, however, they become the same. Hence, it
is also possible to retrieve the photon-number
distribution underlying the revivals through just a single $W_m
(t)$. When the $W_m (t)$ do
not overlap, each of them can be identified with a
single revival of $W(t)$.

We have illustrated the decomposition and also the distribution retrieval for a coherent state and a cat state. For a coherent state,
the retrieval is straightforward. We have also shown how to generate all the
subsequent $W_m (t)$ once $W_0 (t)$ has been determined. As expected,
adding them up yields the full population inversion.
This formalism holds for any state (not just
coherent states), but there may be caveats to consider. For
example, for a cat state, care must be
taking in identifying $W_0(t)$. In this case, the oscillating
distribution leads to additional revivals. We have learned that the additional revivals must not be
thought of as standalone revivals: they are a signature of the high $k$ components of $\chi (k)$ and must be considered with the revivals arising from the low
$k$ components of $\chi (k)$ as part of a single $W_m (t)$ in order for probabilities to be
correctly retrieved. 

The characteristic function approach provides us with a new way of
interpreting the inversion, but photon-number distribution retrieval methods
through single revivals have been known for a while \cite{Fleischhauer:1993dw}. To
highlight the practical advantages of our method, we have at last considered the
case where revivals cannot be resolved anymore. We have outlined
how to retrieve $W_m (t)$ under more general assumptions, and considered a coherent state with average photon-number
of $1$ as an example. By casting aside the limitation imposed by non-overlapping revivals, this work has set the stage for a new
tomographic approach, which will investigate thoroughly in an upcoming work. Other interesting
extensions of this work would be applying the formalism for other
atomic properties, such as the atomic dipole. It would also be worth
looking at how different profiles of $\chi (k)$ lead to
different shapes of $W_m (t)$ by further investigating the propagator
connecting these objects.

\begin{acknowledgments}
The authors would like to thank D. Schmid, K. Marshall, J. Cresswell,
N. Quesada and E. Tham for helpful discussions. This work was supported by NSERC.
\end{acknowledgments}

\bibliography{bibliography.bib}

\end{document}